\documentclass[iop]{emulateapj}
\usepackage{natbib}
\usepackage{threeparttable}
\usepackage{subfigure}
\usepackage{amssymb,amsmath}
\usepackage{hyperref}
\usepackage{graphicx}
\usepackage{epstopdf}
\usepackage{natbib}
\usepackage{rotating} 

\newcommand{\EQ}{\begin{equation}}
\newcommand{\EN}{\end{equation}}
\newcommand{\EQA}{\begin{eqnarray}}
\newcommand{\ENA}{\end{eqnarray}}

\newcommand{\bfom}{\boldsymbol{\omega}}
\newcommand{\bfsij}{\boldsymbol{\mathcal{S}}}

\newcommand{\bfB}{\mathbf{B}}

\newcommand{\uu}{\mbox{\boldmath $u$} {}}

\DeclareMathOperator{\tr}{tr}

\begin{document}
\pagestyle{plain}
\pagenumbering{arabic}

\title{Alignment of the scalar gradient in evolving magnetic fields}
\author{Sharanya Sur\altaffilmark{1}, Liubin Pan\altaffilmark{2}, \& Evan Scannapieco\altaffilmark{1}}
\altaffiltext{1}{School of Earth and Space Exploration, Arizona State University, 
PO Box 876004, Tempe - 85287, USA}
\altaffiltext{2}{Harvard-Smithsonian Center for Astrophysics, 60 Garden St., Cambridge MA 02138, USA}
\email{sharanya.sur@asu.edu, lpan@cfa.harvard.edu, evan.scannapieco@asu.edu}

\begin{abstract}
We conduct simulations of turbulent mixing in the presence of a magnetic field, grown by the small-scale 
dynamo. We show that the scalar gradient field, $\nabla C$, which must be large for diffusion to operate, 
is strongly biased perpendicular to the magnetic field, ${\mathbf B}$. This is true both early-on, when the 
magnetic field is negligible, and at late times, when the field is strong enough to back react on the flow. 
This occurs because $\nabla C$ increases within the plane of a compressive motion, but ${\mathbf B}$ 
increases perpendicular to it. At late times the magnetic field resists compression, making it harder for 
scalar gradients to grow and likely slowing mixing.

\end{abstract}



\section{Introduction}\label{intro}

When pollutants are added to a magnetized, turbulent medium, the motions will stretch the concentration 
field and the magnetic field lines simultaneously. The random stretching of the concentration field causes 
a cascade to smaller scales, which amplifies scalar gradients and leads to homogenization by molecular 
diffusivity \citep{SS00,PS10, PSS13}. Random stretching also rapidly grows a weak magnetic field to 
dynamically important strengths by the action of the small-scale dynamo. In this mechanism, the magnetic 
field growth is initially exponential, until magnetic back reactions resist stretching and folding of the field lines, 
resulting in saturation \citep{Scheko+04, HBD04, BS05,Cho+09, CR09, Fed+11, BSS12, BS13}.   
Understanding mixing in the presence of such evolving magnetic fields is crucial for astrophysical flows in 
diverse environments, as mixing plays a vital role in intergalactic enrichment \citep{Schaye+03, Pichon+03, 
Pieri+06,Scan+06,Becker+09}, thermal conduction in the magnetized intracluster gas \citep{KCS14},
the dispersion of heavy elements in the interstellar medium \citep{PS07}, and pollution of the pristine 
gas in the early Universe \citep{PSS13}. 

In an earlier work \citep{SPS14}, we reported on direct numerical simulations of scalar mixing in the presence 
of an evolving magnetic field, generated by the small-scale turbulent dynamo. One of the important results 
obtained in that study was the fact that scalar mixing is hindered by dynamically important magnetic fields, 
leading to longer mixing time scales. However, some key questions concerning the mechanism by which 
mixing is slowed remained unanswered. Mixing occurs by molecular diffusion in directions of strong 
concentration gradients, gradients that are built up by the same turbulent motions that cause magnetic field 
amplification. Thus an important, open question is the degree to which the directions of  magnetic fields and 
concentration gradients are aligned. In the context of galaxy clusters, \citet{KCS14} found that random 
motions suppress local thermal conduction by aligning the magnetic fields transverse to the local temperature 
gradient. Similarly, how such alignments evolve over the kinematic and saturated phases of magnetic field 
evolution is likely to have strong consequences for the efficiency of fluid mixing as occurs in a wide variety of 
astrophysical environments.

The purpose of this {\em Letter} is to dwell upon these concerns and develop a coherent physical picture
of passive scalar mixing in a turbulent, magnetized medium. To this end, we analyze the data of a subset 
of simulations reported in \citep{SPS14}, where we performed a suite of numerical simulations of this process 
using the publicly available MHD code FLASH \citep{Fryxell00}. For the aims and scope of this contribution, 
we will focus on the alignments of the magnetic field, the vorticity, and the scalar gradient, encompassing 
both the kinematic and nonlinear phases of magnetic field evolution, which as we will show, provides a unique 
perspective on the underlying physics of both turbulent mixing and dynamo action. This Letter is organized 
as follows. In Section 2, we briefly describe our numerical methods and initial conditions. The results obtained 
from our study and an overview of the overall physical picture they suggest are presented in Section 3. 


\section{Numerical Modeling}\label{init}

The details of the numerical setup are described in \citet{SPS14}. In the following, we only highlight the essential 
features of our simulations.
We adopted an isothermal equation of state, a uniform $512^{3}$ grid with periodic 
boundary conditions, and velocities driven by solenoidal modes at about half the scale of the box. Mixing was 
studied by tracking the evolution of a scalar concentration field, $C({\bf x}, t)$, i.e., the local mass fraction of 
pollutants, driven on the same scale as the velocities \citep{PS10, SPS14}, with homogenization most efficient 
in regions in which the scalar gradient, $\nabla C,$ is  large. In all simulations, the initial magnetic field was 
uniform, pointed in the $z$-direction, with plasma beta $\beta \equiv 2p/B^2 = 10^{7}$, where $p$ is the thermal 
pressure and $B$ is the magnetic field strength. 

We primarily examine a non-ideal run with a Mach number $\mathcal{M}=0.3$, Prandtl and Schmidt 
numbers ${\rm Pm} = {\rm Sc}=1,$ and a fluid Reynolds number ${\rm Re} = 1250,$ defined as the ratio of the 
product of the forcing scale of the turbulence and the three-dimensional velocity dispersion to the viscosity.
In addition, we also consider two ideal runs at $\mathcal{M}=0.3$ and $\mathcal{M}=2.4$. 
In these runs, the dissipation of kinetic and magnetic energies as well as the scalar variance is through 
numerical diffusion, and thus the effective magnetic Prandtl and Schmidt numbers are both of $\approx O(1)$. 
We use the unsplit staggered mesh algorithm in FLASHv4 with a constrained transport scheme to maintain 
$\nabla\cdot\bfB$ to machine precision \citep{LD09, Lee13} and the HLLD Riemann solver \citep{MK05}, 
instead of artificial viscosity to capture shocks.
In the context of studying the nature of the alignments, the motivation for including these ideal runs is twofold - to 
explore if the qualitative nature of the alignments is sensitive to the presence or absence of explicit physical 
dissipation and secondly, to study the effect of compressibility. 


\section{Results}\label{results}

\subsection{Scalar Gradients, Magnetic Fields, and Vorticity}

\begin{figure*}[ht!]
\centering
\includegraphics[width=1.75\columnwidth]{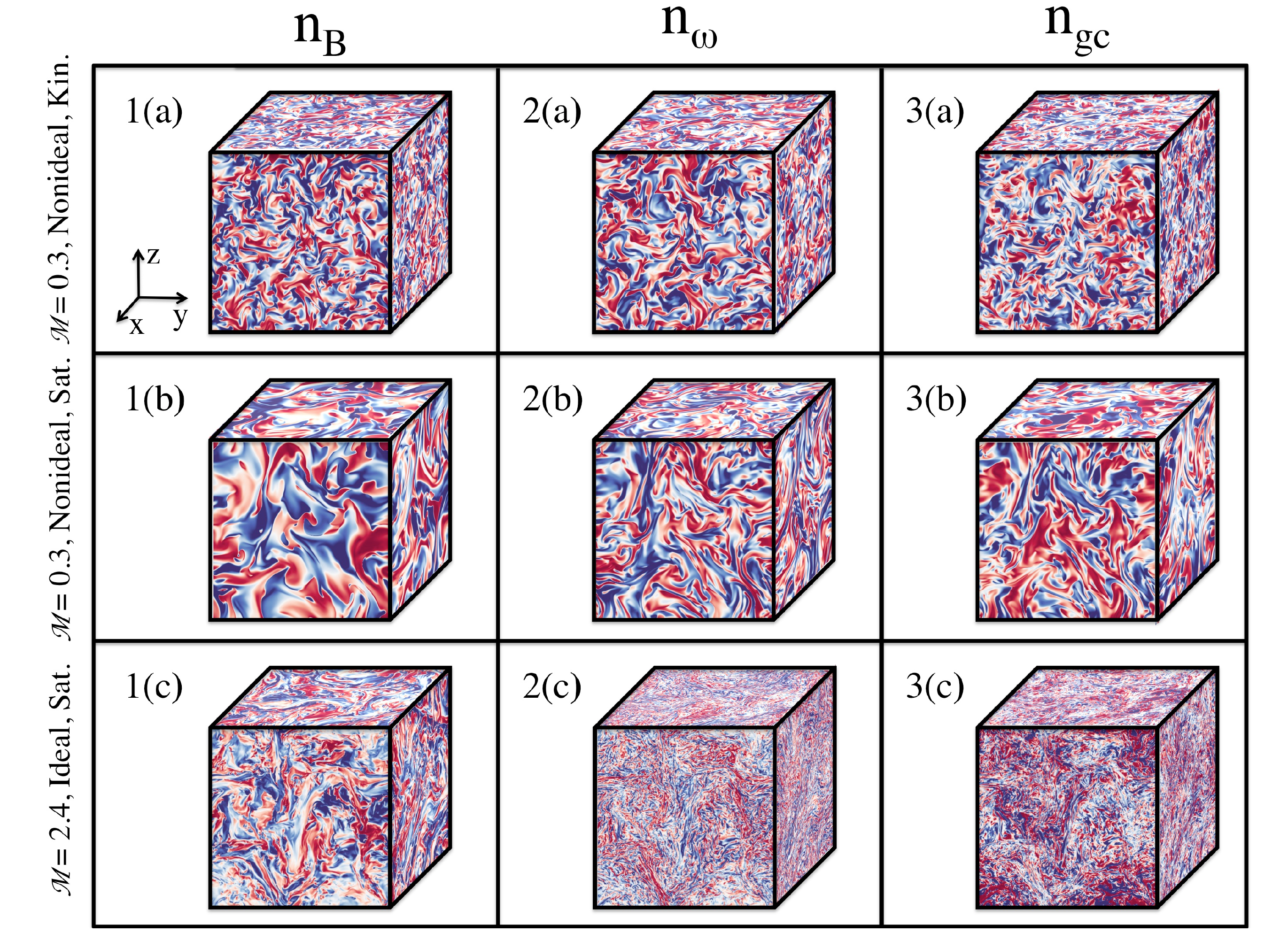}
\caption{Three-dimensional rendering of the projection of unit vectors of the magnetic 
field (${\bf n}_{B}$), the vorticity (${\bf n}_{\omega}$), and the scalar gradient (${\bf n}_{\rm gc}$). 
The first and second rows correspond to the kinematic and saturated phases of the magnetic 
evolution for the non-ideal $\mathcal{M}=0.3$ runs at $t/t_{\rm ed}=2.5$ and $8.8$ respectively. 
The last row shows the saturated phase of the ideal $\mathcal{M}=2.4$ run at $t/t_{\rm ed}=22$. 
Blue denotes vectors pointing outwards, red denotes vectors pointing inwards, and white denotes 
vectors along the faces.}
\label{fig:volrend}
\end{figure*}

In Figure~\ref{fig:volrend}, we show three-dimensional volume renderings of the 
projections of the directional unit vectors of the magnetic field (${\bf n}_{\rm B} \equiv {\bfB}/|B|$), 
the vorticity (${\bf n}_{\omega} \equiv {\bfom}/|\omega|$), and the scalar gradient  
(${\bf n}_{\rm gc} \equiv  {\nabla C}/|\nabla C|$) for the non-ideal $\mathcal{M}=0.3$ and ideal 
$\mathcal{M} = 2.4$ runs. For the subsonic run, we show both the kinematic and saturated phases 
of magnetic field evolution at times $t=2.5\,t_{\rm ed}$ and $8.8\,t_{\rm ed}$ respectively, while 
for the supersonic run we show only the saturated phase at $t=22\,t_{\rm ed}$. Here $t_{\rm ed}$ 
is the eddy-turnover time. The saturated values of the magnetic to kinetic energies are 
$\approx 0.36$ for the subsonic run and $\approx 0.16$ for the supersonic run.
As evident from the figure, in both kinematic and the saturated phases and for both $\mathcal{M}=0.3$ 
and $2.4$, a positive correlation between ${\bf n}_{\rm B}$ and ${\bf n}_{\omega}$ is evident, whereas, 
${\bf n}_{\rm gc}$ appears to align transversely to both ${\bf n}_{\omega}$ and ${\bf n}_{\rm B}$. 

\begin{figure}[ht!]
\centering
\hspace{-1.13cm}
\includegraphics[width=0.9\columnwidth]{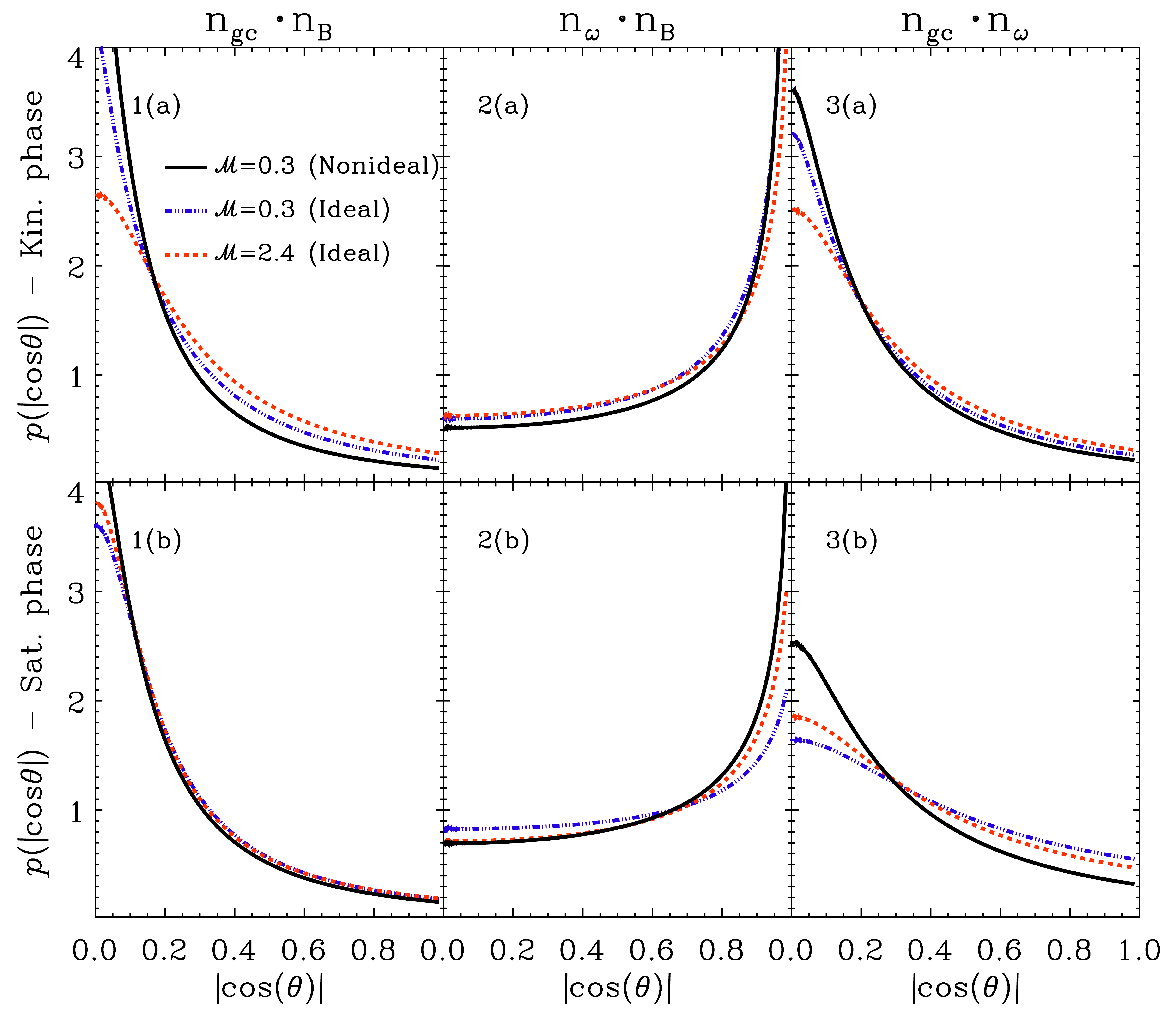}
\caption{PDFs of the alignments of ${\bf n}_{\rm B}$, ${\bf n}_{\rm gc}$ and
${\bf n}_{\omega}$ in the kinematic (upper row) and saturated phase (lower row) of magnetic field 
amplification. The data correspond to the $\mathcal{M}=0.3$ (black solid) and $2.4$ (red dashed) 
ideal runs, and a non-ideal $\mathcal{M}=0.3$ (blue dash-dotted) run. }
\label{fig:align1}
\end{figure}

To quantify these visual impressions, we show in Figure 2 the probability distribution functions (PDFs) 
of the cosines of the angles between the three vectors in all three simulations considered here.
For the two subsonic runs, the PDFs are averaged over $t =(1.5-4)\,t_{\rm ed}$ in the kinematic phase
and over $t=(6-9)\,t_{\rm ed}$ in the saturated phase. For $\mathcal{M}=2.4$, they are averaged over 
$t=(3-9)\,t_{\rm ed}$ and $t=(21-25)\,t_{\rm ed}$ respectively.
In all runs, and in both the kinematic and saturated phases, the direction of the scalar gradient tends to be  
orthogonal to the magnetic field, implying that scalar dissipation occurs primarily perpendicular to ${\bf n}_{B}$. 
The figure also confirms that magnetic field and vorticity are aligned in the two phases. Such a correlation was 
also observed in \cite{BNST95, Miller96}, although the degree of alignment decreases once the magnetic 
field attains saturation. In the kinematic phase, we find that there is a strong tendency of the scalar gradient 
to be perpendicular to the vorticity \cite{Ash87, Kerr87}, which also becomes weaker in the saturated phase. 
Apart from minor differences, it is evident that the alignment of the three unit vectors is qualitatively 
independent of the flow compressibility and the presence or absence of explicit dissipative terms.    

Insight into these alignments can be gained from the evolution equations of ${\bfB}$,
 ${\nabla C},$ and ${\bfom}.$ The magnetic  and scalar fields are deformed by the velocity 
gradient ${\partial u_{i}}/{\partial x_{j}}$,  which can be decomposed into a rate of strain 
tensor $\mathcal{S}_{ij} \equiv ({\partial u_{i}}/{\partial x_{j}} + {\partial u_{j}}/{\partial x_{i}})/2 - (\partial_{k}u_{k})\delta_{ij}/3$,
a rate of expansion tensor, $(\partial_{k}u_{k})\delta_{ij}/3$,
and an antisymmetric tensor $\Omega_{ij} \equiv \epsilon_{ijk} \omega_k/2$, 
corresponding to the vorticity. 
From the magnetic induction and the scalar equations, we have,  
 \EQ
\frac{D {\bfB}}{D t} = {\bfB}\cdot\bfsij - \frac{2}{3}\left(\nabla\cdot \uu\right)\,{\bfB} 
-\frac{1}{2}\left({\bfB}\times\bfom\right) + \eta \nabla^2 {\bfB},
\label{ind1}
\EN
and 
\EQ
\frac{D\nabla C}{D t}=-\nabla C\cdot\bfsij-\frac{1}{3}\left(\nabla\cdot \uu\right)\nabla C
- \frac{1}{2}\left(\nabla C\times\bfom\right) + \kappa \nabla^2(\nabla C),
\label{scalar1}
\EN
where $D/Dt $ is the Lagrangian derivative, and $\eta$ and $\kappa$ are 
the resistivity and diffusivity, respectively. \footnote{For simplicity, we 
ignored the spatial fluctuations of $\eta$, $\kappa$, and the density, which would 
cause extra terms like $(\nabla^2 C)\nabla \kappa$. A similar situation occurs for eq.\ \ref{vort} below.}
In the above equations, the terms proportional to $\bfom$ correspond to the simultaneous rotation 
of ${\bfB}$ and $\nabla C$ by the vorticity, which leave their magnitudes unchanged. However, 
the divergence terms may change the amplitudes of ${\bfB}$ and $\nabla C$, but not their directions. 
Instead, it is the strain that continuously amplifies both ${\bfB}$ and $\nabla C$ and determines their 
relative orientation. Flux-freezing suggests that ${\bfB}$ increases normal to compressive directions of 
the strain, but $\nabla C$ is amplified along the compressive directions. Therefore, the directions of 
${\bfB}$ and $\nabla C$ always tend to be perpendicular, as seen in Figure 2. Such an orthogonal 
orientation in the kinematic phase was also reported by \cite{KCS14}, in the context of thermal conduction 
in galaxy clusters, where it leads to a suppression of the thermal flux.

Ignoring the large-scale driving force, the vorticity equation reads, 
\EQ
\frac{D {\bfom}}{D t} = {\bfom}\cdot\bfsij -\frac{2}{3}\left(\nabla\cdot \uu\right){\bfom}
+\frac{\nabla\rho\times\nabla p}{\rho^{2}} + \nabla\times \boldsymbol{L} + \nu \nabla^2  {\bfom},
\label{vort}
\EN
where $p$ is the thermal pressure, and $\boldsymbol{L} = \boldsymbol{J} \times {\bfB}/\rho$ 
is the acceleration by the Lorentz force with $\boldsymbol{J}= \nabla \times {\bfB}/4\pi$ 
the current density. In our isothermal simulations, the baroclinic term vanishes, and  $\boldsymbol{L}$ 
is negligible in the kinematic phase. 
Similar to the evolution of ${\bfB}$, the divergence term in equation \ref{vort} does not influence the 
direction of ${\bfom}$, but the strain amplifies ${\bfom}$ in the extensive or stretching direction(s). 
Due to angular momentum conservation, a compression by the strain leads to an increase in the 
vorticity in the normal direction \citep{Ash87}. In the saturated phase, the Lorentz effect becomes 
important, causing a weaker alignment of $\bfom$ with $\bfB$ and weaker orthogonal orientation 
with $\nabla C$, as seen in Figure~\ref{fig:align1}. Interestingly, we also find the direction of $\boldsymbol{L}$
is strongly aligned with ${\bf n}_{\rm gc}$ in both the kinematic and saturated phases, indicating that 
one effect of the Lorentz force is to oppose compression. This is related to the buildup of the magnetic 
pressure gradient along the compressive direction(s) of the strain.  

\begin{figure}[t!]
\centering
\includegraphics[width=1.0\columnwidth]{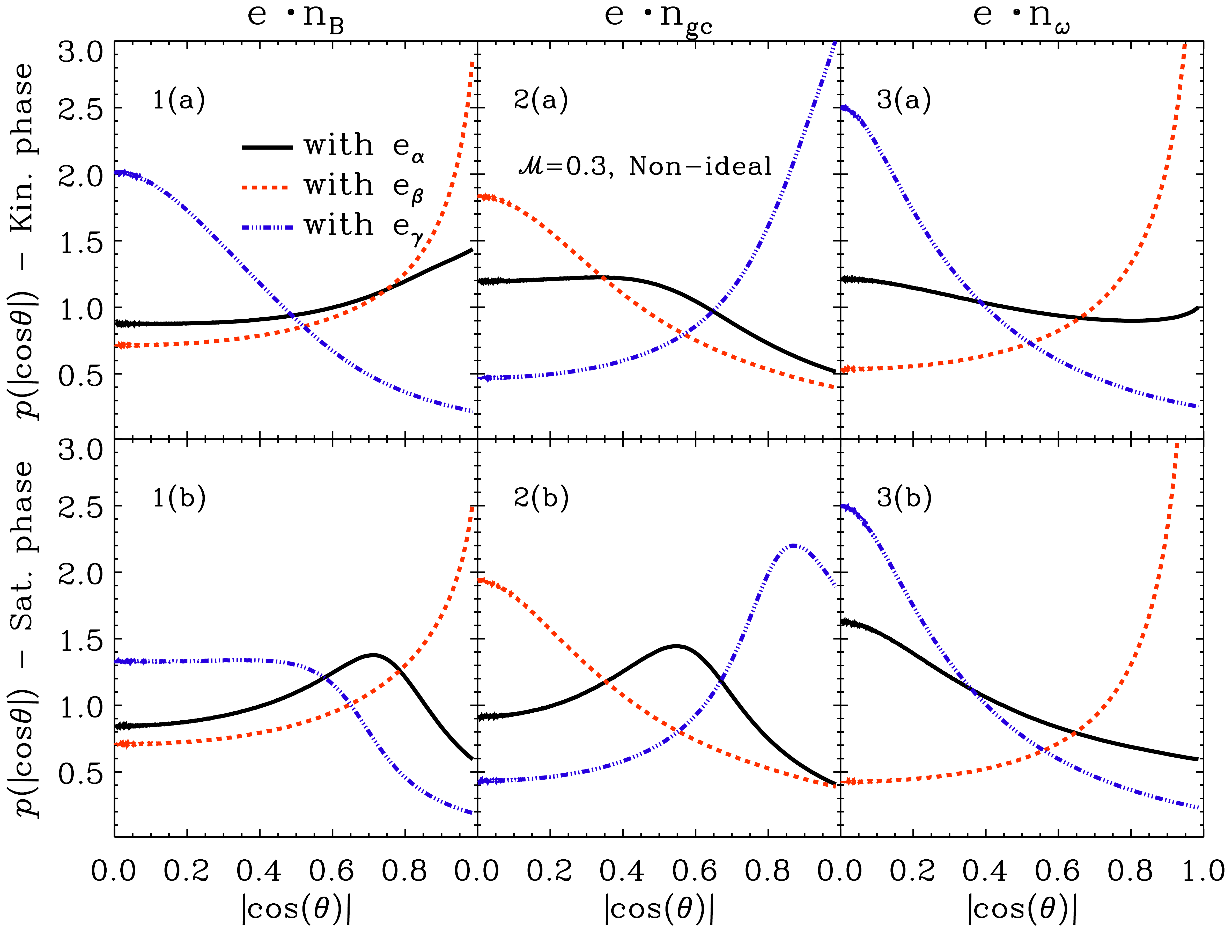}
\caption{PDFs of the alignments of ${\bf n}_{\rm B}$, ${\bf n}_{\rm gc}$ and 
${\bf n}_{\omega}$ with eigenvectors of $\bfsij$ for the non-ideal 
$\mathcal{M}=0.3$ run in the kinematic (upper row) and saturated phase (lower row), 
respectively. Black solid, red dashed, and blue dash-dotted lines denote alignments 
w.r.t to ${\bf e}_{\alpha}$, ${\bf e}_{\beta}$ and ${\bf e}_{\gamma}$, respectively.}
\label{fig:align2}
\end{figure}

\subsection{Strain Tensor}

As the amplifications of $\bfB$, $\bfom,$ and $\nabla C$ all depend on the strain tensor, ${\bfsij}$, 
an analysis of their orientations with respect to the eigenvectors of ${\bfsij}$ sheds further physical 
insight into the problem. We denote the eigenvectors of ${\bfsij}$ by ${\bf e}_{\alpha}, {\bf e}_{\beta}$, 
and ${\bf e}_{\gamma}$, corresponding to the eigenvalues $\alpha > \beta >\gamma$, respectively, with
$\alpha+\beta+\gamma=0$. Thus, ${\bf e}_{\alpha}$ and ${\bf e}_{\gamma}$ 
correspond to the directions of stretching and compression respectively, and the intermediate 
eigenvector ${\bf e}_{\beta}$ has been found to be generally extensive \citep{Ash87}. 

The evolution of the eigenvalues and eigendirections can be derived from the equation 
for the strain tensor,  
\EQA
\frac{D {\bfsij} }{D t}&=&-{\bfsij} \cdot {\bfsij}+\frac{{\bfsij} : {\bfsij}}{3}\boldsymbol{\mathcal{I}}
- \frac{1}{4}(\boldsymbol{\omega}\boldsymbol{\omega}-\frac{\omega^2} {3}\boldsymbol{\mathcal{I}}) \nonumber \\
&-& \frac{2}{3} (\nabla \cdot \boldsymbol{u}){\bfsij} - \boldsymbol{\mathcal{P}} 
+\frac{\tr(\boldsymbol{\mathcal{P}})}{3}\boldsymbol{\mathcal{I}}+\boldsymbol{\mathcal{L}}, 
 \label{strain}
\ENA 
where $\boldsymbol{\mathcal{I}}$ is the unit tensor, $\boldsymbol{\mathcal{P}} = {\nabla (\nabla p }/{\rho})$ 
is the thermal pressure hessian, 
$\mathcal{L}_{ij} = ({\partial L_{i}}/{\partial x_{j}} + {\partial L_{j}}/{\partial x_{i}})/2 - \partial_k L_k \delta_{ij}/3$, and
the viscous term is omitted for simplicity. The terms proportional to $\boldsymbol{\mathcal{I}}$ ensure that $\bfsij$ 
is traceless while the vorticity term is related to a centrifugal effect that tends to produce an expansion perpendicular 
to $\bfom$. The  $- 2( \nabla \cdot \boldsymbol{u} ) {\bfsij}/3$ term suggests that local compressions (expansions) 
by compressible modes in a supersonic flow may amplify (reduce) the amplitude of ${\bfsij}$. 
In the coordinate system of local eigenvectors, equation \ref{strain} suggests that the eigenvalues of $\bfsij$ 
are affected by the diagonal components of the tensors on the right hand side, while the off-diagonal 
components cause the rotation of the eigenvectors.  
For example, a calculation of ${\bf e}_\alpha \cdot D\mathcal{S}/Dt \cdot {\bf e}_\beta$ using equation \ \ref{strain} 
yields, 
\begin{equation}
\frac{D {\bf e}_\alpha }{D t} \cdot   {\bf e}_\beta = -\frac{1}{\alpha-\beta} \left(\frac{\omega_\alpha \omega_\beta}{4} 
+   \mathcal{P}_{\alpha \beta}  - \mathcal{L}_{\alpha\beta}\right), 
\label{rotation}
\end{equation}
where ${D {\bf e}_\alpha }/{D t} \cdot   {\bf e}_\beta$ corresponds to the  projection of the directional change of 
${\bf e}_\alpha$ onto the $ {\bf e}_\beta$ direction, $\mathcal{P}_{\alpha\beta} = {\bf e}_\alpha \cdot \mathcal{P} \cdot {\bf e}_\beta$ 
and $\mathcal{L}_{\alpha\beta} = {\bf e}_\alpha \cdot \mathcal{L} \cdot {\bf e}_\beta$ \citep{NP98}. 
Similar equations can be derived for ${D {\bf e}_\beta}/{D t} \cdot   {\bf e}_\gamma$ and 
${D{\bf e}_\gamma }/{D t} \cdot   {\bf e}_\alpha$. 

In Figure~\ref{fig:align2}, we plot the PDFs of the cosines of the angles 
between ${\bf n}_{\rm B}, {\bf n}_{\rm gc}, {\bf n}_{\omega},$ and the three 
eigenvectors of $\bfsij$ for the non-ideal $\mathcal{M}=0.3$ flow. 
Similar results were obtained for the ideal runs with $\mathcal{M}=0.3$ and $2.4$. 
This similarity suggests that the compressible modes do not significantly affect the relative 
directions of the vectors considered here. While it is clear that the divergence term leaves the 
directions of ${\bf n}_{\rm B}$, ${\bf n}_{\rm gc}$, and ${\bf n}_{\omega}$ unchanged, quite remarkably, 
it also has no direct role in influencing the eigendirections of ${\bfsij},$ as seen from equation~\ref{rotation}. 
However, compressible modes could lead to {\it indirect} effects on the eigendirections,
due to changes in the local amplitudes of the pressure, $\bfB$, $\bfom,$ and the strain eigenvalues. 
That the alignments remain qualitatively unchanged suggests that such effects are minor. 
Furthermore, the fact that the PDFs obtained from the ideal $\mathcal{M}=0.3$ are qualitatively 
similar to the non-ideal run implies that molecular dissipation does not play a direct role in 
determining the qualitative nature of the alignments discussed here. However, a full quantitative 
prediction for the statistics of various angles would require an examination of the dissipation rates 
of $\bfB$, $\bfom$ and $\nabla C$ in each eigen direction of $\bfsij$ following the method of \citet{NP98}. 

In the kinematic phase, the orientations of ${\bf n}_{\rm gc}$ and ${\bf n}_{\omega}$ are 
consistent with previous results for hydrodynamical flows \citep{Kerr87, Ash87, Sree97}. 
The scalar gradient is strongly aligned with ${\bf e}_{\gamma}$, the only compressive direction 
of the strain, and thus orthogonal to the directions in which stretching occurs. The vorticity 
lies preferentially in the ${\bf e}_{\alpha}$-${\bf e}_\beta$ plane, where the strain is extensive, but 
is more strongly-aligned with the intermediate eigenvector ${\bf e}_{\beta}$ rather than 
the principal eigenvector, ${\bf e}_{\alpha}$ \citep{Kerr87, Ash87}. 
Although vorticity is produced primarily along ${\bf e}_{\alpha}$, the angle between ${\bf n}_{\omega}$ and 
${\bf e}_{\alpha}$ shows a rather uniform distribution, because vorticity is converted to the direction 
${\bf e}_{\beta}$ due to a rotation of the eigenvectors by the off-diagonal components  
in equation~\ref{rotation} \citep{NP98}. For example, the $\omega_{\alpha}\omega_{\beta}$ term 
causes a rotation of ${\bf e}_{\alpha}$ and ${\bf e}_{\beta}$ that leads to a preferential alignment 
of $\bfom$ with ${\bf e}_{\beta}$ \citep{SJO91}.  
Finally, the direction of ${\bfB}$ relative to the eigenvectors is similar to $\bfom$ due to the similarity 
of their equations in the kinematic phase, although a difference exists in their PDFs with ${\bf e}_{\alpha}$, 
which may be caused by the vorticity term in equation~\ref{ind1} for $\bfB$. The fact that the 
magnetic field in the kinematic phase lies in the ${\bf e}_{\alpha}$-${\bf e}_\beta$ 
plane dates back to the earlier analytic studies of \citet{Zeldo+84, CFKV99, SC07}. 

In the saturated phase, the alignments of ${\bf n}_{\rm B}, {\bf n}_{\rm gc},$ and ${\bf n}_{\omega},$ 
with ${\bf e}_{\beta}$ are qualitatively similar to those in the kinematic phase, but there are notable 
changes in the PDFs with the other two eigenvectors, especially for  ${\bf n}_{B}$ and ${\bf n}_{\rm gc}$. 
The peak of the PDF for ${\bf n}_{B}$ with ${\bf e}_{\alpha}$ shifts from $0^{\circ}$ to $\approx 45^{\circ}$
and the peak for ${\bf n}_{B}$ with ${\bf e}_{\gamma}$ shifts from $90^{\circ}$ to
$\approx 55^{\circ},$ as also reported in \citep{BNST95, Bran95}.  
Not measured before is the shift of the peak of the PDF of ${\bf n}_{\rm gc}$ with ${\bf e}_{\alpha}$ 
from $90^{\circ}$ to $\approx 60^{\circ}$ and that of ${\bf n}_{\rm gc}$ with ${\bf e}_{\gamma}$ from 
$0^{\circ}$ to $\approx 30^{\circ}$. These changes suggest a rotation of 
${\bf e}_{\alpha}$ and ${\bf e}_{\gamma}$ by ${\mathcal L},$ which can occur through the off-diagonal 
component, ${\mathcal L}_{\alpha\gamma}$, in an equation similar to equation\ \ref{rotation}. As $\boldsymbol {L}$ 
does not directly act on $\bfB$ and $\nabla C$, it  would not affect the tendency of 
${\bf n}_{\rm gc}$ to align perpendicularly to ${\bf n}_{\rm B}$ (Figure\ 2). 

Unlike ${\bf n}_{\rm B}$ and ${\bf n}_{\rm gc}$, no peak shift is observed in the PDFs of the ${\bf n}_{\omega}$ 
relative to the eigenvectors, except for a slightly stronger anti-correlation  with ${\bf e}_{\alpha}$. Note that 
unlike $\bfB$ and $\nabla C$, the Lorentz force acts directly on the vorticity (equation\ \ref{vort}), and the direction 
change of $\bfom$ by $\nabla \times \boldsymbol{L}$ appears to roughly compensate the rotation of the 
eigenvectors by the off-diagonal components of $\mathcal{L}$. Thus, as the magnetic strength increases 
and saturates, the alignment of ${\bf n}_{\omega}$ with the eigenvectors remains largely unchanged. 
A consequence of $\boldsymbol{L}$ directly acting on $\bfom$ but not on $\bfB$ and $\nabla C$ 
is a weaker alignment of ${\bf n}_\omega$ with $\bf n_{\rm B}$ and a weaker transverse correlation
of ${\bf n}_\omega$ with $\bf n_{\rm gc}$ in the saturated phase. 

The weaker alignment of $\nabla C$ with ${\bf e}_{\gamma}$ in the saturated phase has important 
consequences for mixing, as it leads to a slower amplification of the scalar gradient. The Lorentz force 
also opposes the compressive motion of the strain, and decreases the amplitude of the compressive 
eigenvalue,  which further decrease the mixing rate. In fact, in simulations with an
isotropic $\kappa,$ mixing was found to be suppressed in the saturated phase \cite{SPS14}.
Furthermore, the orthogonality of $\bfB$ and $\nabla C$ at all times implies that in a medium with 
anisotropic $\kappa$, mixing is likely to be suppressed even in the kinematic phase.

\subsection{Overview}

Taken together, the features of the various alignments described 
here lead to the following coherent physical picture. For simplicity, we consider the incompressible 
case in which a compressive motion is in the ${\bf \hat x}$ direction and the extensive motions occur 
in the ${\bf \hat y}$-${\bf \hat z}$ plane. In the kinematic phase, this will - (i) increase the concentration 
gradient in ${\bf \hat x}$ and decrease it in the other directions, (ii) increase the magnetic field in 
${\bf \hat y}$ and ${\bf \hat z}$, and decrease it in ${\bf \hat x}$ because of flux freezing, and (iii) increase 
the vorticity in ${\bf \hat y}$ and ${\bf \hat z}$, and decrease it in ${\bf \hat z}$, because of angular 
momentum conservation. In the saturated phase, a substantial Lorentz force, $\boldsymbol{L}$, develops, 
which rotates $\bfom$ and the eigenvectors of the strain. The directions of $\bfB$ and $\nabla C$ are 
not directly affected by $\boldsymbol{L}$, and therefore they remain largely orthogonal. However, the 
Lorentz force tends to be aligned with $\nabla C$, thereby resisting the compressive motion of the strain, 
leading to slower mixing. More detailed analysis of the Lorentz force and its effect on $\bfom$ and the 
strain tensor, and the dissipation rates of ${\bfB}$, $\nabla C$, and ${\bfom}$ in the eigendirections of 
$\bfsij$  will provide further insight into the coevolution of magnetic field growth and scalar mixing 
in the presence of turbulence.\\

SS \& ES were supported by the National Science Foundation under grant AST11-03608 and NASA 
theory grant NNX09AD106. LP acknowledges support from the Clay postdoctoral fellowship at 
Harvard-Smithsonian Center for Astrophysics. The authors would also like to acknowledge the Advanced 
Computing Center at Arizona State University (URL: http://a2c2.asu.edu/), the Texas Advanced Computing 
Center (TACC) at The University of Texas at Austin (URL: http://www.tacc.utexas.edu), and  the Extreme 
Science and Engineering Discovery Environment (XSEDE) for providing HPC resources via grant 
TG-AST130021 that have contributed to the results reported within this paper. The FLASH 
code is developed in part by the DOE-supported Alliances Center for Astrophysical Thermonuclear 
Flashes (ASC) at the University of Chicago.

\end{document}